\documentclass[runningheads]{llncs}
\usepackage{listings}

\usepackage{color}
\usepackage{cite}
\usepackage[dvipsnames]{xcolor}
\usepackage{amsmath,amsfonts}
\usepackage{multicol}
\usepackage{enumerate}
\usepackage{mathtools}
\usepackage{mathabx}
\usepackage{url}
\usepackage{comment}

\usepackage{subcaption}
\usepackage{algorithm}
\usepackage[noend]{algpseudocode}

\usepackage{graphicx}

\usepackage{tikz}
\usetikzlibrary{shapes.gates.logic.US,trees,positioning,arrows}
\usetikzlibrary{calc, positioning}

\usepackage{forest}

\usetikzlibrary{graphs}
\usetikzlibrary{arrows,automata}

\forestset{%
  angle below/.style={
    tikz+={%
      \draw ($(!1.child anchor)!.25!(.parent anchor)$) [bend right=15] to ($(.parent anchor)!.75!(!l.child anchor)$);
    },
  },
  arrow below/.style={
    tikz+={%
      \draw[->] ($(!1.child anchor)!.30!(.parent anchor)$) [bend right=20] to ($(.parent anchor)!.7!(!l.child anchor)$);
    },
  },
}
\newcommand{\OG}[1]{{\color{blue} OG: #1}}

\newcommand{\et}{\textit{et al.}}

\newcommand{\OR}{\texttt{OR}}
\newcommand{\AND}{\texttt{AND}}
\newcommand{\SAND}{\texttt{SAND}}

\newcommand{\state}{{s}}
\newcommand{\solnstate}[1]{{\state_{s#1}}}
\newcommand{\startstate}{{\state_0}}

\newcommand{\action}{{Action}}
\newcommand{\actions}{{Actions}}

\newcommand{\actionitem}{{b}}
\newcommand{\actionset}{{\mathbb{B}}}
\newcommand{\OperatorSet}{{\{\OR,\AND,\SAND\}}}
\newcommand{\operator}{{\Delta}}

\newcommand{\Top}{{top}}
\newcommand{\Rad}{{rad}}
\newcommand{\Kid}{{kid}}

\newcommand{\StateSet}{{S}}
\newcommand{\EdgeSet}{{F}}
\newcommand{\EdgeFunc}{{f}}
\newcommand{\EdgeMap}[1]{{\EdgeFunc:#1}}
\newcommand{\StartStateSet}{{S_0}}
\newcommand{\EndStateSet}{{S_s}}

\newcommand{\PStateSet}{{P}}
\newcommand{\QStateSet}{{Q}}

\begin{document}
%
\title{A Novel Approach for Attack Tree to Attack Graph Transformation: Extended Version}
%
\author{Nathan D. Schiele\inst{1}\orcidID{0000-0003-1186-1503} \and
Olga Gadyatskaya\inst{1}\orcidID{0000-0002-3760-9165} }
%
%
\institute{Leiden Institute of Advanced Computer Science, Leiden University, The Netherlands \\
\email{\{n.d.schiele, o.gadyatskaya\}@liacs.liedenuniv.nl}}
%
\maketitle              
\begin{abstract}
Attack trees and attack graphs are both common graphical threat models used by organizations to better understand possible cybersecurity threats. These models have been primarily seen as separate entities, to be used and researched in entirely different contexts, but recently there has emerged a new interest in combining the strengths of these models and in transforming models from one notation into the other. The existing works in this area focus on transforming attack graphs into attack trees. In this paper, we propose an approach to transform attack trees into attack graphs based on the fundamental understanding of how actions are represented in both structures. From this, we hope to enable more versatility in both structures.

\end{abstract}

\section{Introduction}

\emph{Attack trees} are a common and useful tool for threat modeling. They allow us to present attack components in a graphical structure that is relatively easily explained and understood. Each node in an attack tree represents a action, and its children represent actions in service to their parent action. The relationship between the children of a node describe the relationship between the components, \OR, \AND\ and \SAND, describing if all components or only one component need to be completed, and in which order. The major advantage of this model is its compactness, allowing for even complex threats to be modeled concisely. However, attack trees have several downsides, particularly when understanding how the attacker interacts with the system. For instance, it is not evident from an attack tree what is a possible plan of a successful attack in this system nor is it simple to discuss overall mitigation strategies.

An alternative modeling structure, which addresses these downsides, are \emph{attack graphs}~\cite{SheynerAttackGraph2002}. Attack graphs represent all possible states a system may hold, and the transitions between those states. The major disadvantage of attack graphs is their size, with even small system models resulting in excessively large attack graphs \cite{valmariStateExplosionProblem1998}.
Attack graphs are not as common in the security industry as the succinct nature of attack trees makes them more appealing to human experts. Additionally, attack graphs are arguably less intuitive, inhibiting their adoption and use by non-technical security experts \cite{haqueEvolutionaryApproachAttackSurvey}.

 However, attack graphs and attack trees represent similar information, and both are valid approaches for modeling potential attack vectors.
Additionally, for both attack graphs and attack trees, there is a major consideration of how these models are to be generated, with automated generation at the forefront of current threat model research~\cite{widel2019beyond,barik2016attack,kaynar2016taxonomy}. 

These two threat models each have their disadvantages, and these disadvantages are seemingly offset, or are easier to offset, by using the other model. It would be useful to have a way to transform one model into the other, and thus be able to combine their perspectives. However, to the best of our knowledge, there is currently no well-defined transformation between these two models. There have been works proposing methods of converting an attack graph into an attack tree~\cite{pinchinatSynthesisAttackTrees2014,haqueEvolutionaryApproachAttack2017,hongScalableAttackRepresentation2013}. Yet, to date, there has been little work into transforming an attack tree into an attack graph. This is the problem that we address in the present paper.


\section{Related Work}

Attack trees (ATs) have been fairly widely studied to date. First introduced by Schneier in 1999 as an efficient and effective means of conveying attack information~\cite{schneier1999attack}; several researchers have worked to develop the threat model further. Mauw and Oostdijk developed propositional and multiset semantics for attack trees~\cite{mauwAttackTrees2006}. Jhawar \et\ developed a refinement of the sibling relationships, adding a Sequential AND or ``SAND'' relation, alongside discussing the utility and semantic implications (the SP graph semantics) of such a refinement~\cite{jhawarAttackTreesSequential2015}. Many works have attempted to generate attack trees automatically, given that many attack trees in industry are generated manually, and effective automatic attack tree generation would be a valuable contribution to this space
\cite{hongScalableAttackRepresentation2013,vigoAutomatedGenerationAttack2014,
ivanovaTransformingGraphicalSystem2016,gadyatskayaRefinementAwareGenerationAttack2017}.

Attack graphs (AGs) similarly have been widely studied. Sheyner \et\ laid out a formal, syntactic structure of a state-based attack graph and proposed a model checking method to generate attack graphs~\cite{SheynerAttackGraph2002}. Noel and Jajodia focused on using topological aggregation to manage the complexity of attack graphs~\cite{noelEfficientMinimumCostNetwork2003,noelManagingAttackGraph2004}.
Ou \et\ described a more scalable methodology for generating attack graphs from a logical model~\cite{ouScalableApproachAttack2006}. In a broad review of previous works on attack graphs, Lippmann and Ingols found that scalability was the biggest limiting factor in the overall use of attack graphs, with attack graphs quickly becoming too large to be useful~\cite{lippmannAnnotatedReviewPapers2005}. Other problems with attack graphs according to~\cite{lippmannAnnotatedReviewPapers2005} include prescribing meaning to the states within the attack graph structure and the usability of attack graphs as a communication tool; given their size and complexity, attack graphs can be hard to use in industry, as developing recommendations from attack graphs is often too difficult. The primary recommendation in~\cite{lippmannAnnotatedReviewPapers2005} was that future work on attack graphs would need to pay special attention to scalability. 

An important topic in the attack graph literature is the meaning behind attack graph nodes. They most frequently represent the states of the associated system after some actions are taken. However, as Lallie \et\ points out, attack graphs in literature suffer from inconsistency, and state definitions and labelling schema are not above these inconsistencies \cite{lallie2020review}. In our review of the literature, we have found much of the literature can be broadly classified into two categories in the context of the meaning or labeling of attack graph states, those that use logical conditions roughly based on state as in \cite{ouScalableApproachAttack2006}, or those that derive state meaning from some underlying system as in \cite{pinchinatSynthesisAttackTrees2014}. We take our cues from the literature in that final group, our states will be defined based on a set of actions defined from a system model. We note that other types of attach graphs assign different meaning to states, e.g, vulnerabilities or system hosts~\cite{kaynar2016taxonomy}.

There have been some works on conversion between AGs and ATs, with the research largely focused on the conversion from attack graphs to attack trees.  

\paragraph{AG $\rightarrow$ AT.}
Pinchinat \textit{et al.}, focused on generating attack trees from attack graphs using the ATSyRA approach~\cite{pinchinatSynthesisAttackTrees2014,pinchinatATSyRaIntegratedEnvironment2016}. We are attempting the reverse transformation, and a similar designed library based methodology also from Pinchinat \et\ of these transformation give us insight into how such transformations can function~\cite{pinchinatSynthesisAttackTrees2014}. 

Dawkins and Hale, focused on an analysis of network models, and in that analysis described a method of creating attack chains representing complete or near complete attack vectors~\cite{dawkinsSystematicApproachMultiStage2004}. These attack chains could then be used to construct attack trees. The primary purpose of the work was not to develop a transformation between attack graphs and attack trees, and thus the transition is not fully developed~\cite{dawkinsSystematicApproachMultiStage2004}. Most recently, Haque and Atkinson have overviewed existing approaches to generate attack graphs and attack trees, and to convert attack graphs into attack trees, and have found that those works have suffered from inefficiency or inaccuracy~\cite{haqueEvolutionaryApproachAttackSurvey}. 


Hong \et\ describe that previous attempts to define a transformation from attack graphs to attack trees have failed due to the exponential size increases of attack graphs~\cite{hongScalableAttackRepresentation2013}. The state explosion problem broadly affects any state based modeling system, of which attack graphs are one; or better put by Valmari, "the number of states of almost any system of interest is huge". The biggest issue with the state explosion problem is the computational complexity, as the number of states in a system increases exponentially, the difficulty of handling this information becomes far more complex, as well as time and space inefficient. There are a number of common strategies used to minimize and mitigate the \emph{state explosion problem} (SEP)~\cite{valmariStateExplosionProblem1998}, many of which we follow as discussed in Section~\ref{sec:discussion}. A major concern for any attack graph generation scheme will be handling the exponential number of states~\cite{ouScalableApproachAttack2006}, and given a transformation from attack trees to attack graphs is a form of attack graph generation in itself, addressing the state explosion problem will be a major concern for us as well.

\paragraph{AT $\rightarrow$ AG.}
To the best of our knowledge, there have been no works on transforming attack trees into attack graphs. The reason for this gap could be that attack trees are more succinct threat models, not suffering from the state explosion problem, and are thus more handy for security analysis with human experts~\cite{dawkinsSystematicApproachMultiStage2004}. However, creating a transformation into attack trees opens up new interesting research directions, for example, new approaches to automatically generate attack graphs starting from automatically generated attack trees, or using existing, human-designed attack trees to capture security issues in relevant systems with application of the attack graph-based security monitoring. 

Finally, attack trees and attack graphs have been very popular graphical security notations in the last 20 years, being featured in many research papers. We refer the interested reader to several surveys for more details about these notations and their applications: on attack trees~\cite{widel2019beyond} and on attack graphs~\cite{barik2016attack,feng2011survey,zeng2019survey}, and on both attack trees and attack graphs, possibly among other graphical security models,~\cite{lallie2020review,kordy2014dag,hong2017survey}.

\section{Definitions}
\label{sec:defs}

While many formal definitions of both attack trees (ATs) and attack graphs (AGs) exist, we will start from the following definitions. These definitions were selected because they share common properties with many modifications of both attack trees and attack graphs, and thus should enable modification of the algorithms described below to enable further development of this methodology. Specifically, we use a recursive attack tree definition \emph{{\`a} la} Gadyatskaya \et~\cite{gadyatskayaRefinementAwareGenerationAttack2017}, and a basic state-based attack graph  definition from Sheyner \et~\cite{SheynerAttackGraph2002}.

\begin{definition}[Attack Tree]
  \label{def:AT}
Let $\actionset$ denote a set of actions, \OR\ and \AND\ be two unranked associate and commutative operators that are disjunctive and conjunctive respectively, and \SAND\ be an unranked associate but non-commutative conjunctive operator. An attack tree $t$ is an expression over $\actionset \cup \OperatorSet$ generated by the following formal grammar (for $\actionitem \in \actionset$):
\[
t ::= \actionitem | \actionitem \triangleleft \OR(t,\dots,t)| \actionitem \triangleleft \AND(t,\dots,t) | \actionitem \triangleleft \SAND(t,\dots,t)
\]

\end{definition}

This definition is recursive, as each subtree is a complete attack tree in and of itself. A single action $\actionitem$ by itself is an attack tree, and this fact will later be used to help develop our mapping schema in Section~\ref{sec:mapping}.

\begin{definition}[Attack Graph]
  \label{def:AG}
An attack graph or $AG$ is a tuple $(\StateSet, \EdgeFunc, \StartStateSet, \EndStateSet)$, where $\StateSet$ is a set of states, $\EdgeMap{ \StateSet \times \actionset \rightarrow \StateSet}$ is a partial function that defines the transition relation for $\StateSet$ by the set of actions $\actionset$, $\StartStateSet \subseteq \StateSet$ is a set of initial states and $\EndStateSet \subseteq \StateSet$ is a set of success states.
\end{definition}

This definition of attack graphs is the same proposed by Pinchinat \et~\cite{pinchinatSynthesisAttackTrees2014}, who defined an attack graph to attack tree transformation, the inverse of what we propose. The $\actionset$ is a set of actions based in the system model the attack graph is built around. Ultimately, the contents of $\actionset$ will be wholly defined by such a system model. By convention, we define partial function with a set of mappings, given in the set $\EdgeSet$.
Expanding upon Definition~\ref{def:AT}, we can define two further functions that will be useful in the transformation algorithm. Both of these functions are inspired by the work of Gadyatskaya \et~\cite{gadyatskayaRefinementAwareGenerationAttack2017} and focus on isolating specific elements of attack trees.

\begin{definition}[Top Function]
  \label{def:func:top}
This function obtains the action of the root node as follows (for $\operator \in \OperatorSet$):
\[
\Top(\actionitem) = \Top(\actionitem \triangleleft \operator(t_1,\hdots,t_n)) = \actionitem
\]
\end{definition}

We call an attack tree of depth one a \emph{radical}. The auxiliary function $\Rad$ defined below obtains a single transition (radical) from one level of an attack tree to another.
\begin{definition}[Rad Function]
  \label{def:func:rad}
The $\Rad$ function finds the uppermost radical of a provided tree as input as follows (for $\operator \in \OperatorSet$):
\[
\Rad(t) = \Rad(\actionitem \triangleleft \operator(t_1,\hdots,t_n)) = \actionitem \triangleleft \operator(\Top(t_1),\hdots,\Top(t_n))
\]
\end{definition}

The previous definition introduces the idea of radical elements of attack trees. The intuition behind these radicals is that a \emph{radical} in an attack tree is the smallest possible subtree. It is a component consisting solely of a root node, an operator and a set of children that are singular nodes themselves.  Our attack graph transformation will use radicals as base components. As outlined in the definition above, a radical is found by using the \textit{rad} function. The child nodes in a radical can themselves be root nodes of different radicals, and these radicals are likewise found by using the \textit{rad} function on those child nodes.

\begin{definition}[Kid Function]
  \label{def:func:kid}
The $\Kid$ function returns the set of children of a radical:
\[
\Kid(t) = \Kid(\actionitem \triangleleft \operator(t_1,\hdots,t_n)) = \{\Top(t_1),\hdots,\Top(t_n)\}
\]
\end{definition}

Fundamentally, our transformation approach will take the form of defining the separate radicals, creating a mapped transformation of those radicals from an attack tree to attack graph, and then combining the radicals that are in the attack graph in a manner that retains the attack component information expressed in attack trees.

\paragraph{Zero and Single Element Radicals.}

The radicals described thus far are of size $n$; they are dynamically defined such that any number of children in the radical can be directly mapped from attack trees to attack graphs. The implication is that $n \ge 2$, however this is not necessary. There are two edge cases that are worth mentioning, when $n=0$ and when $n=1$. When $n=0$, we have the case of a radical without children; this is the case of $\Rad(\actionitem) = \actionitem$. Additionally, we have a graphical example of this single action mapping in Figure~\ref{fig:sig_node_ex}.

When $n=1$, this would fit our definition of a radical, as there would be a defined root, operator and set of children with cardinality 1.
However, our three defined radicals all converge when $n=1$.
This follows from our understanding of these operators. The difference between \AND\ and \SAND\ being that \SAND\ is an ordered \AND; when $n=1$, there is only one possible order, thus \AND\ and \SAND\ are equivalent. For similar reasons, \OR\ is also the same as both \AND\ and \SAND.

We introduce these edge cases for the sake of completeness of our approach. However, single element radicals are rare in real-world attack trees as their children nodes are often considered redundant and removed from the tree.

\section{Transformation Example}\label{sec:example}

\begin{figure}[b!]
\centering
\scalebox{.8}{
\scriptsize
\begin{forest}
  for tree={
    draw,
    minimum height=1cm,
    anchor=parent,
    align=center,
    child anchor=parent,
    edge=<-
  },
  adnode/.style={rounded rectangle,},
  [{Get Root Access}, adnode
    [{Exploit Buffer Overflow}, adnode
      [{Deploy \texttt{.rhhost} file}, adnode],
      [{Remote login}, adnode]],
    [{Exploit Administrator}, adnode, angle below
      [{Invent Need For \\Root Access}, adnode]
      [{Befriend \\Administrator}, adnode, arrow below
        [{Get Phone Number}, adnode],
        [{Invite to Social Function}, adnode]
      ]
    ]
  ]
\end{forest}}
\caption{A simple attack tree}
\label{fig:AttackTree}
\end{figure}
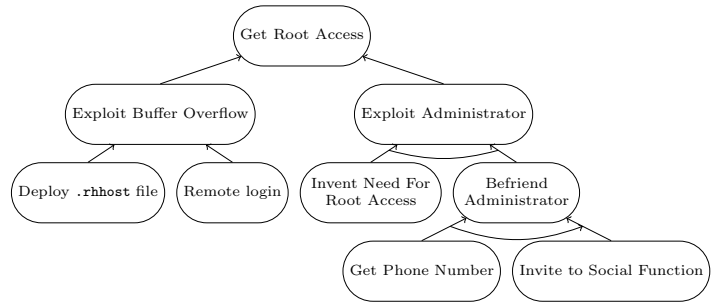

\def\yone{.7}
\def\ytwo{2.2}
\def\nd{3}
\def\xone{.5}
\def\xtwo{3.5}
\def\fy{1.6}

\begin{figure}[t!]
\centering
\scalebox{.8}{
\scriptsize
\begin{tikzpicture}[->,>=stealth',shorten >=1pt,auto, node distance=\nd cm,
                    semithick]
  \tikzstyle{every state}=[fill=white,draw=black,text=black]

  \node[state] (0)                    {$\state_{0}$};
  \node[state]         (2) [below left = \fy and \xone of 0] {$\state_{2}$};
  \node[state]         (1) [left of=2] {$\state_{1}$};
  \node[state]         (3) [below right of=0] {$\state_{3}$};
  \node[state]         (4) [right of=3]  {$\state_{4}$};
  \node[state]         (5) [below  =\yone cm of 1]  {$\state_{5}$};
  \node[state]         (6) [below  =\yone cm of 2]  {$\state_{6}$};
  \node[state]         (7) [below = \yone cm of 5]  {$\state_{7}$};
  \node[state]         (8) [below = \yone cm of 6]  {$\state_{8}$};
  \node[state]         (9) [below =\ytwo of 3]  {$\state_{9}$};
  \node[state]         (10) [below =\ytwo of 4]  {$\state_{10}$};
  \node[state]         (11) [below right = \yone cm and 1 cm of 7]  {$\state_{11}$};
  \node[state]         (12)  [below = \yone cm of 11] {$\state_{12}$};
  \node[state]         (S1) [below = \yone cm of 12]  {$\state_{s1}$};
  \node[state]         (S2) [below =\ytwo of 9]  {$\state_{s2}$};
  \node[state]         (S3) [below =\ytwo of 10]  {$\state_{s3}$};

  \path (0) edge   [bend right]            node [above]{Get Phone Number} (1)
            edge              node [left, pos=.4, align=left]{Invent Need \\for Root Access} (2)
            edge              node {Remote Login} (3)
            edge   [bend left]           node {Deploy \texttt{.rhhost} file} (4)
        (1) edge node [pos=.4, align=left]{Invite to\\Social Function} (5)
        (2) edge node [pos=.4, align=left]{Get phone\\number} (6)
        (3) edge node [pos=.4, align=left]{Exploit Buffer \\Overflow} (9)
        (4) edge node [pos=.4, align=left]{Exploit Buffer \\Overflow} (10)
        (5) edge node [pos=.4, align=left]{Befriend\\Administrator}  (7)
        (6) edge node [pos=.4, align=left]{Invite to \\Social Function}  (8)
        (7) edge node [pos=.4, align=left]{Invent Need\\for Root Access}  (11)
        (8) edge node [pos=.4, align=left]{Befriend\\Administrator}  (11)
        (11) edge node [pos=.4, align=left]{Exploit Buffer \\Overflow}  (12)
        (9) edge node []{Get Root Access} (S2)
        (10) edge node []{Get Root Access} (S3)
        (12) edge node []{Get Root Access} (S1);

\end{tikzpicture}
}
\caption{An attack graph representation of the model in Figure \ref{fig:AttackTree}}
\label{fig:AttackGraph}
\end{figure}
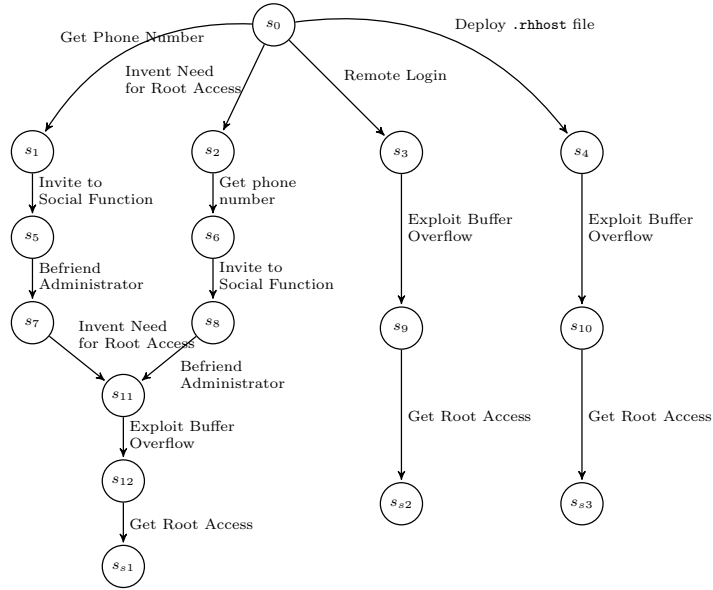

We begin with a simple example of an attack tree as seen in Figure \ref{fig:AttackTree}. The overall goal is to get root access of a system. Directly below the overall root, we see two sub goals in an \OR\ relationship; if any sub-goal, either exploiting a buffer overflow or exploiting an administrator, is accomplished, then the overall goal of getting root access will be accomplished. With exploiting the buffer overflow, we can see that there are two attack components, again in an \OR\ relationship. Once again, if either component is completed, then the sub-goal is accomplished and by extension the overall goal is accomplished. With exploiting the administrator, we see two attack components in an \AND\ relationship, meaning that both components will need to be accomplished to accomplish the sub-goal. One of these components has subcomponents in a \SAND\ relationship, which need to be accomplished in a particular order before the parent action can be completed.

In the provided example, the order of subcomponents for exploiting the administrator is not important. There are \AND\ relationships where the order of components is important or where the components specifically need to occur in parallel~\cite{jhawarAttackTreesSequential2015}; however, we are limiting our understanding of \AND\ relationships to be unordered \AND\ relationships, i.e., the attacker can execute actions in any order to successfully achieve the goal. The \SAND\ component of the attack tree is an ordered \AND, where a defined order is provided. The elements in the \SAND\ relationship occur in the same order (first the administrator phone number needs to be obtained, then the administrator can be invited somewhere).


In Figure \ref{fig:AttackGraph}, we see an attack graph representation of the same information. First, we see that the nodes of the AT have now become the state transitions in the AG. We start from an initial state, with the only description of this state being that no attack component has been applied yet. The initial attack vectors (transitions outgoing from the initial state) are the basic components in the AT (the leaf nodes). We see some basic structures that will enable us to generalize the transformation procedure. Namely, we see that the components in an \OR\ relationship (``Remote login'' and ``Deploy \texttt{.rhhost} file'') create fairly parallel paths through the attack graph. Additionally, for the \AND\ relationship (``Invent need for root access'' and ``Befriend Administrator''), we see a type of a lattice structure, where transitions between states occur internally within the \AND\ components before returning to an overall path. The generalization of these rules is introduced in the following section.

\section{Mapping}
\label{sec:mapping}

Fundamentally, nodes in attack trees represent actions, while nodes in attack graphs represent system states. The edges in an attack tree represent the relationships between actions, while the edges in attack graphs represent state transitions. Our intuition is that state transitions and actions are equivalent concepts, and as such, we will base our transformation on this equivalence. We now present our transformation approach for radicals.

We create a distinction of $\PStateSet$ and $\QStateSet$ states, where $\PStateSet$ states are already in the attack graph, while $\QStateSet$ states are generated by a specific mapping.

\subsection{Edge Cases}

\paragraph{Single node.}
To begin, let us consider the basic case of an attack tree with a single action or node, and what this extreme case would be in the form of an attack graph.

\begin{figure}[t!]
\centering
\scalebox{.8}{
\begin{subfigure}{.5\textwidth}
   \centering
  \vspace{1cm}
  \begin{forest}
    for tree={
      draw,
      minimum height=1cm,
      anchor=parent,
      align=center,
      child anchor=parent,
      edge=<-
    },
    adnode/.style={rounded rectangle,},
      [{Action}, adnode, ]
  \end{forest}
  \vspace{1cm}
  \caption{Attack Tree}
  \label{sfig:AT_sig_node_ex}
\end{subfigure}
\begin{subfigure}{.5\textwidth}
   \centering
  \begin{tikzpicture}[->,>=stealth',shorten >=1pt,auto, node distance=2cm,
                      semithick]
    \tikzstyle{every state}=[fill=white,draw=black,text=black]

    \node[state,dashed]         (0)  {$\startstate$};
    \node[state,dashed]         (1) [below  of=0] {$\solnstate{}$};

    \path (0) edge              node [right]{Action} (1);

  \end{tikzpicture}
  \caption{Attack Graph}
  \label{sfig:AG_sig_node_ex}
\end{subfigure}
}
\caption{Single action attack tree and attack graph}
\label{fig:sig_node_ex}
\end{figure}
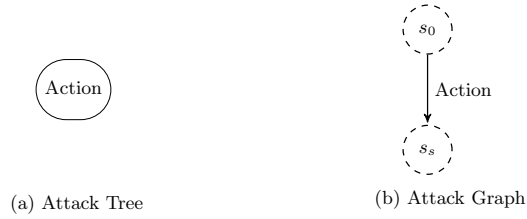

From Figure~\ref{fig:sig_node_ex}, we can see that the attack tree of a single action is very simple, it is merely a single node named after the action. In the attack graph, we have a similar mechanism to represent this action; however, this mechanism is in the form of a state transition, otherwise referred to as an edge. In graph theory, for an edge to exist, there must be nodes for an edge to exist between \cite{biggs1986graph}; similarly, in attack graphs, we have a tuple that containing states and state transitions, and state transitions cannot exist without states.

We see our first example of $\PStateSet$ and $\QStateSet$ states. In the case of the single node attack tree in Figure~\ref{fig:sig_node_ex}, only a state transformation is described; however, as we just discussed, a state transition implies the existence of states in the attack graph. In order to resolve this paradox where state transitions need to exist without being created directly, we instead say that the defined state transition goes between two $\PStateSet$ states. These $\PStateSet$ were already in the attack graph, and thus do not need to be created for this mapping. Following from Definition~\ref{def:AG}, the complete form of this simple attack graph is given in Equation~\ref{eq:sig_node_ex}.
\begin{equation}
\footnotesize
\begin{aligned}\label{eq:sig_node_ex}
\PStateSet &= \{\startstate, \solnstate{}\} \\
\QStateSet &= \emptyset \\
\EdgeSet &=\{(\startstate,\text{\action}) \mapsto \solnstate{}\}\\
AG &= \left(\PStateSet\cup\QStateSet, \EdgeMap{\EdgeSet}, \{\startstate\}, \{\solnstate{}\}\right)
\end{aligned}
\end{equation}
Where the only two states, $\startstate$ and $\solnstate{}$ are the starting and solution states respectively, and the action is the defined state transition between them.

\paragraph{Radical with one child.}

\begin{figure}[b!]
\centering
\scalebox{.8}{
\begin{subfigure}{.5\textwidth}
   \centering
  \begin{forest}
    for tree={
      draw,
      minimum height=1cm,
      anchor=parent,
      align=center,
      child anchor=parent,
      edge=<-
    },
    adnode/.style={rounded rectangle,},
      [{Action 1}, adnode,
      [{Action 2}, adnode]
      ]
  \end{forest}
  \caption{Attack Tree}
  \label{sfig:AT_sig_rad_ex}
\end{subfigure}
\begin{subfigure}{.5\textwidth}
   \centering
  \begin{tikzpicture}[->,>=stealth',shorten >=1pt,auto, node distance=1.5cm,
                      semithick]
    \tikzstyle{every state}=[fill=white,draw=black,text=black]

    \node[state,dashed]         (0)  {$\startstate$};
    \node[state]         (1) [below  of=0] {$s_1$};
    \node[state,dashed]         (2) [below  of=1] {$\solnstate{}$};

    \path (0) edge              node [right]{Action 2} (1)
          (1) edge              node [right]{Action 1} (2);

  \end{tikzpicture}
  \caption{Attack Graph}
  \label{sfig:AG_sig_rad_ex}
\end{subfigure}
}
\caption{Single child attack tree and attack graph}
\label{fig:sig_rad_ex}
\end{figure}
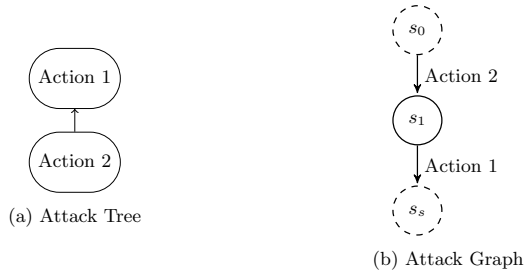

From Figure \ref{fig:sig_rad_ex}, we see a similar structure in the attack graph as in Figure \ref{fig:sig_node_ex}. Both actions in the attack tree are directly mapped to state transitions in the attack graph. However, unlike in Figure \ref{fig:sig_node_ex}, we now have two edges, and as all edges require a defined starting and ending state, we will require a third node in the attack graph.
\cite{biggs1986graph,SheynerAttackGraph2002}. As such, these two edges necessitate the creation of an intermediate state. The meaning behind this state is it is the resulting state after the completion of ``\action\ 2'' but before the completion of ``\action\ 1''. This is our first encounter with $\QStateSet$ states. As in Equation~\ref{eq:sig_node_ex}, we have the same starting $\PStateSet$ states; however, now we introduce a $\QStateSet$ state, as the generated intermediate state for this mapping. The $\QStateSet$ state was not already in the attack graph. The resulting attack graph would be of the form:
\begin{equation}
\footnotesize
\begin{aligned}\label{eq:sig_rad_ex}
\PStateSet &= \{\startstate, \solnstate{}\} \\
\QStateSet &= \{\state_1\} \\
\EdgeSet &=\{(\startstate,\text{\action\ 2}) \mapsto \state_{1},  (\state_1,\text{\action\ 1})\mapsto \solnstate{}\}\\
AG &= \left(\PStateSet\cup\QStateSet, \EdgeMap{\EdgeSet}, \{\startstate\}, \{\solnstate{}\}\right)
\end{aligned}
\end{equation}


 We can begin to develop an intuition regarding how our transformation will be structured. Namely, nodes in attack trees will be directly mapped to edges (the state transitions) in attack graphs, and states will be created to enable such state transitions to exist. The states are given meaning by where they fall in the transition relation partial function. The only remaining concept that requires mapping is the operators (\AND, \SAND, and \OR) in attack trees. Once the operator mapping is defined, we will have the means to create a transformation algorithm from attack trees to attack graphs.

\vspace{-8pt}
\subsection{OR Radical}
\label{ssec:OR}

Our understanding of the \OR\ radical in attack trees is that only a single component (child action) of an \OR\ radical needs to be completed for the goal, or root, of the \OR\ to also be accomplished.
In an attack graph where individual attack vectors are expressed more explicitly, we would expect that each individual \OR\ component would contribute to a separate attack vector. These separate attack vectors will introduce separate states in the attack graph, as each attack component is a different state transition, and different state transitions result in different states.

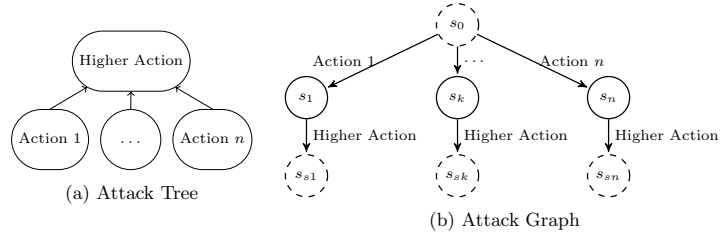
\begin{figure}[t!]
\centering
\scalebox{.8}{
\scriptsize
\begin{subfigure}{.4\textwidth}
  \centering
  \begin{forest}
    for tree={
      draw,
      minimum height=1cm,
      anchor=parent,
      align=center,
      child anchor=parent,
      edge=<-
    },
    adnode/.style={rounded rectangle,},
      [{Higher \action}, adnode, 
        [{\action\ 1}, adnode]
        [{$\hdots$}, adnode]
        [{\action\ $n$}, adnode]
      ]
  \end{forest}
\caption{Attack Tree}
\label{fig:OR_AT_rad}
\end{subfigure}
\begin{subfigure}{.6\textwidth}
  \centering
  \begin{tikzpicture}[->,>=stealth',shorten >=1pt,auto, node distance=1.3cm,
                      semithick]
    \tikzstyle{every state}=[fill=white,draw=black,text=black]

    \node[state,dashed] (0) [below left of=0] {$\startstate$};
    \node[state]         (1) [below left =.7cm and 2cm of 0] {$\state_{1}$};
    \node[state]         (n) [below right =.7cm and 2cm of 0] {$\state_{n}$};
    \node[state]         (2) at ($(1)!0.5!(n)$) {$\state_{k}$};
    \node[state,dashed]         (5) [below of=1]  {$\solnstate{1}$};
    \node[state,dashed]         (8) [below  of=n]  {$\solnstate{n}$};
    \node[state,dashed]         (7) at ($(5)!0.5!(8)$)  {$\solnstate{k}$};

    \path (0) edge              node [left]{\action\ 1} (1)
              edge              node {$\hdots$} (2)
              edge              node [right]{\action\ $n$} (n)
          (n) edge node [pos=.5]{Higher \action} (8)
          (2) edge node [pos=.5]{Higher \action} (7)
          (1) edge node [pos=.5]{Higher \action} (5);

  \end{tikzpicture}
\caption{Attack Graph}
\label{fig:OR_AG_rad}
\end{subfigure}
}
\caption{\OR\ Radical}
\label{fig:OR_rad}
\end{figure}

If we express solely this radical (that is, $rad(t) = t$) as an attack graph, we would generate the following graph. By convention, we organize separate the transition of different levels of the attack graph into separate $\EdgeSet$ sets and combine them for the final definition of the partial function transition relation $\EdgeFunc$. As with Equation~\ref{eq:sig_rad_ex}, the $\PStateSet$ states are already in the attack tree, while $\QStateSet$ states are generated from this radical:

\begin{equation}
\footnotesize
\begin{aligned}\label{eq:OR}
\PStateSet &= \{\startstate,\solnstate{1}, \hdots, \solnstate{n}\} \\
\QStateSet &= \{\state_1,\hdots,\state_n\} \\
\EdgeSet_1 &=  \{(\startstate,\text{\action\ 1}) \mapsto \state_1, \hdots, (\startstate,\text{\action\ n}) \mapsto \state_n\}\\
\EdgeSet_2 &=\{(\state_1,\text{Higher \action}) \mapsto \solnstate{1} , \hdots, (\startstate,\text{Higher \action}) \mapsto \solnstate{n}\}\\
AG &= \left(\PStateSet\cup\QStateSet, \EdgeMap{\EdgeSet_1\cup\EdgeSet_2},  \{\startstate\}, \{\solnstate{1},\hdots,\solnstate{n}\}\right)
\end{aligned}
\end{equation}

As we can see in Figures~\ref{fig:OR_AT_rad}~and~\ref{fig:OR_AG_rad}, and subsequently in Equation~\ref{eq:OR}, the \OR\ radical in an attack graph is a series of disjoint ``legs'' with a state transition culminating in a state for each attack component represented in the attack tree. As each attack component is unique, the system states that are a result of the application of those attack components must also be unique. This does not rule out the possibility of the application of unique attack components resulting in the same state; that would be a possible expansion of the radical mapping we have laid out here. Given that the ``Higher \action'' in this radical could itself be a component in another radical, we see that the ``Higher \action'' in the attack tree radical is again represented as a state transition in the attack graph. Here, we assume that only one component in the OR relationship will ever be completed in a single attack vector.

 Given that the \actions\ $1$ through $n$ in Figure~\ref{fig:OR_AG_rad} are distinct state transitions, we expect that the resulting states are different.
The subgraph below the \OR\ radical is thus duplicated for each element in the \OR\ radical. If the only remainder of the graph below the \OR\ radical was a solution state, as in the case of $rad(t) = t$, then there would be a separate solution state for each element in the \OR\ radical. If we consider the transition relation $\EdgeFunc$ and the example of Figure~\ref{fig:OR_AG_rad}, we operate under the expectation $f(\state_1, \text{\action\ 1}) \ne f(\state_n, \text{\action\ 1})$; our expectation is the output of these functions are entirely different states in $AG$. This does not preclude the possibility of the output of different $\EdgeFunc$ inputs resulting in the same state, merely that we expect that with different state inputs, the output states would thus be different. Intuitively, if the same action is applied to two different states, the resulting state would be different.

\subsection{AND Radical}
\label{ssec:AND}

On the contrary to the \OR\ operator, \AND\ operators have an understanding of all actions in the \AND\ needing to be performed for the \AND\ to be completed. However, unlike the \SAND\ operator, there is no defined order to the actions in an \AND. Thus, the attack graph must represent all possible combinations of action orders for the actions within a single \AND\ operator.

\begin{figure}[b!]
\centering
\scalebox{.8}{
\scriptsize
\begin{subfigure}[t]{.4\textwidth}
  \centering
  \begin{forest}
    for tree={
      draw,
      minimum height=1cm,
      anchor=parent,
      align=center,
      child anchor=parent,
      edge=<-
    },
    adnode/.style={rounded rectangle,},
      [{Higher \action}, adnode, angle below
        [{\action\ 1}, adnode]
        [{\action\ 2}, adnode]
        [{\action\ 3}, adnode]
      ]
  \end{forest}

\caption{AND Radical}
\label{fig:AND_AT_rad}
\end{subfigure}

\begin{subfigure}[t]{.6\textwidth}
  \centering
  \begin{tikzpicture}[->,>=stealth',shorten >=1pt,auto, node distance=2cm,
                      semithick]
    \tikzstyle{every state}=[fill=white,draw=black,text=black]

    \node[state,dashed] (0) [below left of=0] {$\startstate$};
    \node[state]         (1) [below left=  .5cm and 2.5cm of 0] {$\state_1$};
    \node[state]         (3) [below right=.5cm and 2.5cm of 0] {$\state_3$};
    \node[state]         (2) at ($(1)!0.5!(3)$) {$\state_2$};
    \node[state]         (4) [below = .5cm  of 1] {$\state_4$};
    \node[state]         (6) [below= .5cm of 3] {$\state_6$};
    \node[state]         (5) at ($(4)!0.5!(6)$) {$\state_5$};
    \node[state]         (7) [below = .5cm of 5] {$\state_7$};
    \node[state,dashed]         (G) [below = .3cm of 7 ] {$\state_8$};

    \path (0) edge              node [sloped]{\scriptsize \action\ 1} (1)
              edge              node {\scriptsize\action\ 2} (2)
              edge              node [sloped]{\scriptsize\action\ 3} (3)
          (1) edge              node [left]{\scriptsize\action\ 2} (4)
              edge              node [sloped, pos=.2]{\scriptsize\action\ 3} (5)
          (2) edge              node [sloped,pos=.3] {\scriptsize\action\ 1} (4)
              edge              node [sloped,pos=.3] {\scriptsize\action\ 3} (6)
          (3) edge              node [sloped,pos=.2]{\scriptsize\action\ 1} (5)
              edge              node [right]{\scriptsize\action\ 2} (6)
          (4) edge              node [sloped]{\scriptsize\action\ 3}(7)
          (5) edge              node {\scriptsize\action\ 2}(7)
          (6) edge              node [sloped, pos=.4]{\scriptsize\action\ 1}(7)
          (7)edge              node {\scriptsize Higher \action} (G);

  \end{tikzpicture}
\caption{\AND\ Radical in an Attack Graph}
\label{fig:AND_AG_rad}
\end{subfigure}
}
\caption{\AND\ Radical}
\label{fig:AND_rad}
\end{figure}

Consider the \AND\ radical in Figure~\ref{fig:AND_AT_rad}. Transforming it into an attack graph, we would obtain the following graph (we continue with the $\EdgeSet$ convention presented in Equation~\ref{eq:OR}):

\begin{equation}
\footnotesize
\begin{aligned}\label{eq:AND}
  \PStateSet &= \{\startstate,\state_{n}\} \\
  \QStateSet &= \{\state_1,\hdots,\state_7\} \\
\EdgeSet_1 &=  \{(\startstate,\text{\action\ 1}) \mapsto \state_1, (\startstate,\text{\action\ 2}) \mapsto \state_2,  (\startstate,\text{\action\ 3}) \mapsto\state_3\}\\
\EdgeSet_2 &= \{(\state_1,\text{\action\ 2}) \mapsto \state_4, (\state_1,\text{\action\ 3}) \mapsto \state_5,  (\state_2,\text{\action\ 1}) \mapsto \state_4,\\
        &\text{ }\text{ }\text{ }\text{ }\text{ }\text{ }
            (\state_2,\text{\action\ 3}) \mapsto \state_6,  (\state_3,\text{\action\ 1}) \mapsto \state_5,  (\state_3,\text{\action\ 2}) \mapsto\state_6\}\\
\EdgeSet_3 &=\{ (\state_4,\text{\action\ 3}) \mapsto \state_7,  (\state_5,\text{\action\ 2}) \mapsto \state_7,  (\state_6,\text{\action\ 1}) \mapsto \state_7\}\\
\EdgeSet_4 &=\{ (\state_7,\text{Higher \action}) \mapsto \solnstate{}\}\\
AG &= \left(\PStateSet\cup\QStateSet, \EdgeMap{\EdgeSet_1\cup\EdgeSet_2\cup\EdgeSet_3\cup\EdgeSet_4}, \{\startstate\}, \{\solnstate{}\}\right)
\end{aligned}
\end{equation}

As we can see in Figure~\ref{fig:AND_AG_rad} and subsequently in Equation~\ref{eq:AND}, the \AND\ radical in an attack graph creates a lattice structure such that intermediate states represent the application of some combination of different elements in an \AND\ operator.

Generally speaking, on the $i^\text{th}$ row of the lattice, we see the application of $i$ distinct state transitions. In the first row, every state is the result of a single state transition. On the second, every state is the result of two state transitions. On the $n^\text{th}$ row, every state is the application of $n$ state transitions. We can find the number of states in a row $k$ by calculating the number of unique unordered grouping of $k$ state transitions with $n$ possible state transitions, this is simply ${n\choose k}$. In the final row of the lattice, we have ${n\choose n}$ states, and thus we only have a single state. This follows from our understanding of the meaning of these states, as this state is the application of all the different actions in the \AND\ radical. There is only one possible, unordered, way to apply all the actions in the \AND\ radical, and this is the final resulting state. From this state, we apply the ``Higher \action'', but as this is only applied to a single state at the end of the lattice, we only have a single final state from this radical.

In Figure~\ref{fig:AND_AG_rad}, we see an \AND\ radical with $n=3$, a so-called $3-$\AND. For the lack of space, generalization of the \AND\ is provided in Appendix~\ref{app:sand}. 


\subsection{SAND Radical}
\label{ssec:SSAND}
The \SAND, or \texttt{S}equential \AND, radical is similar to the \AND\ in that all actions in the \SAND\ need to be completed for the operator to evaluate as successful. However, unlike the \AND\ operator, the \SAND\ operator has a specified order. \actions\ in the \SAND\ need to occur sequentially for the overall radical to be successful. For the tree shown in Figure~\ref{fig:SAND_AT_rad}  we would generate the following graph: 

\begin{figure}[t!]
\centering
\scalebox{.8}{
\scriptsize
  \begin{subfigure}[t]{.5\textwidth}
  \centering
  \begin{forest}
    for tree={
      draw,
      minimum height=1cm,
      anchor=parent,
      align=center,
      child anchor=parent,
      edge=<-
    },
    adnode/.style={rounded rectangle,},
      [{Higher \action}, adnode, arrow below
        [{\action\ 1}, adnode]
        [{$\hdots$}, adnode]
        [{\action\ $n$}, adnode]
      ]
  \end{forest}
\caption{Attack Tree}
\label{fig:SAND_AT_rad}
\end{subfigure}

\begin{subfigure}[t]{.5\textwidth}
  \centering
  \begin{tikzpicture}[->,>=stealth',shorten >=1pt,auto, node distance=1.7cm,
                      semithick]
    \tikzstyle{every state}=[fill=white,draw=black,text=black]

    \node[state,dashed] (0) [below left of=0] {};
    \node[state, draw = none]          (1) [below left of=0] {};
    \node[state, draw = none]          (3) [below right of=0] {};
    \node[state]                       (2) at ($(1)!0.5!(3)$) {$S_1$};

    \node[state, draw = none]          (4) [below left of=2]  {};
    \node[state, draw = none]          (6) [below right of=2]  {};
    \node[state]                       (5) at ($(4)!0.5!(6)$)  {$S_k$};

    \node[state, draw = none]          (7) [below left of=5]  {};
    \node[state, draw = none]          (9) [below right of=5]  {};
    \node[state]                       (8) at ($(7)!0.5!(9)$)  {$S_n$};

    \node[state, draw = none]          (10) [below left of=8]  {};
    \node[state, draw = none]          (12) [below right of=8]  {};
    \node[state,dashed]          (11) at ($(10)!0.5!(12)$)  {};

    \path (0) edge              node {\action\ 1} (2)
          (2) edge              node {$\hdots$} (5)
          (5) edge              node {\action\ $n$} (8)
          (8) edge              node {Higher \action} (11);

  \end{tikzpicture}
\caption{Attack Graph}
\label{fig:SAND_AG_rad}
\end{subfigure}
}
\caption{SAND Radical}
\label{fig:SAND_rad}
\end{figure}
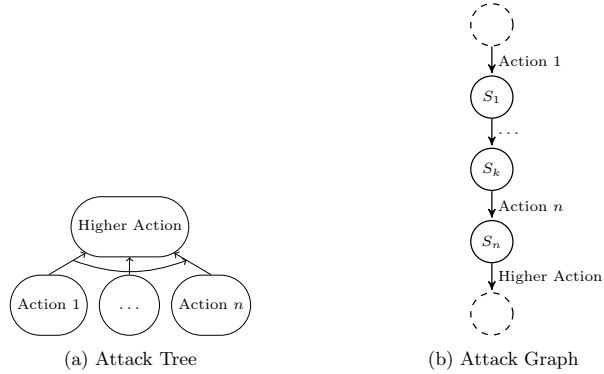

\begin{equation}
\footnotesize
\begin{aligned}\label{eq:SAND}
  \PStateSet &= \{\startstate,\solnstate{}\} \\
  \QStateSet &= \{\state_1,\hdots,\state_n\} \\
\EdgeSet_1 &=  \{(\startstate,\text{\action\ 1}) \mapsto \state_1\}\\
\vdots \\
\EdgeSet_{n} &=\{ (\state_{n-1},\text{\action\ n} ) \mapsto \state_n\}\\
\EdgeSet_{n+1} &=\{ (\state_n,\text{Higher \action}) \mapsto \solnstate{}\}\\
AG &= \left(\PStateSet\cup\QStateSet, \EdgeMap{\EdgeSet_1\cup\hdots\cup\EdgeSet_{n+1}}, \{\startstate\}, \{\solnstate{}\}\right)
\end{aligned}
\end{equation}
As we can see in Figure \ref{fig:SAND_AG_rad}, the \SAND\ radical is by far the simplest structure we have developed thus far. Like the other radicals, the overall starting and ending states are implied, and may be the overall attack graph starting and ending states, or may be internal states created by other radicals. As the order is defined, the actions are chained together in order, with the intermediate states representing the state of the system after the application of some of the actions.

In the \SAND\ radical, actions are applied sequentially. As such, in the attack graph state transitions are also applied sequentially. This results in $n$ overall states (one state for each of $n$ actions). 

\section{Algorithm}
\label{sec:algorithm}

The algorithm is effectively divided in two parts: analysis of the attack tree, and sequential construction of the attack graph. In the first portion of the algorithm, we find all the radicals present in the attack tree and store them in a radical dictionary. This radical dictionary, denoted as $RD$, is a dictionary data structure that represents a deconstructed attack tree, where all elements of the original attack tree are present, and stored separately as radicals. We store each radical using the radical root node value as the key. Once all the radicals in the attack tree are found and stored into the radical dictionary, the attack graph can be systematically constructed. This part is presented in Algorithm~\ref{alg:ATtoAGHard}.

\begin{algorithm}[t!]
\scriptsize
\caption{AT $\rightarrow$ AG Transformation}
\label{alg:ATtoAGHard}
\begin{algorithmic}
\Require Attack Tree, $AT$
\Require Radical Dictionary, $RD$
\For{$\actionitem \in \actionset$}
\State $r \leftarrow \Rad(\actionitem)$
\If{$|\Kid(r)| \ge 0$}
\State Add $\Top(r) : r$ to $RD$
\EndIf
\EndFor
\State Create Attack Graph, $AG = (\{\state_0, \state_s\}, \EdgeMap{\{(\state_0, \Top(AT))\mapsto \state_s)\}}, \{\state_0\}, \{\state_s\})$
\While{$|RD| > 0$}
\For{$e \in AG.\EdgeSet$}
\If{$e$ in Key Set of $RD$}
\State $r \leftarrow RD(e)$
\State \texttt{Add Radical to Attack Graph}$(r, AG)$
\State Remove $r$ from $AG$
\EndIf
\EndFor
\EndWhile
\State Assign values to states in attack graph
\State\Return Attack Graph
\end{algorithmic}
\end{algorithm}

\begin{algorithm}[t!]
\floatname{algorithm}{Procedure}
\scriptsize
\caption{Add Radical to Attack Graph}
\label{alg:RepAGEdge}
\begin{algorithmic}
\Require Attack Tree Radical: $t = \actionitem \triangleleft \operator(c_1,\hdots,c_n), \Rad(t) = t$
\Require Attack Graph: $AG = (\StateSet, \EdgeMap{\EdgeSet}, \StartStateSet, \EndStateSet)$
\State Remove $(\state_j,\actionitem) \mapsto \state_k$ from $\EdgeSet$
\State $AG_k \leftarrow$ the subgraph where $\state_k \in AG_{k}.\StartStateSet$
\If{$\operator = \OR$}
\State Remove $AG_k$ from $AG$
\For{$c_i$ in $\Kid(t)$}
\State $AG_{k+i} \leftarrow AG_k$
\State Add $\state_{j+i}$ to $\QStateSet$ and $(\state_j, c_i) \mapsto \state_{j+i}$ to $\EdgeSet$
\State Add $\state_{j+ n +i}$ to $\QStateSet$ and $(\state_j, c_i) \mapsto \state_{j+n+i}$ to $\EdgeSet$
\State Add $AG_{k+i}.\StateSet$ to $\QStateSet$ and $AG_{k+i}.\EdgeSet$ to $\EdgeSet$
\State Add $(\state_{j+n+i}, c_k) \mapsto \state_{k+i}$, where $\state_{k+i} \in AG_{k+i}.\StartStateSet$ to $\EdgeSet$
\EndFor
\ElsIf{$\operator = \AND$}
\State Create subgraph $AG_k = (\QStateSet, \EdgeMap{\EdgeSet_k},  \emptyset, \emptyset)$
\State $\QStateSet\leftarrow \{\state_{1},\hdots,\state_{n},\state_{n+1}, \hdots, \state_{2^n-1}\} $
\State $\EdgeSet_1 \leftarrow \{(\state_{1},c_2 ) \mapsto \state_{n+1},\hdots, (\state_{n},c_{n-1}) \mapsto \state_{n\choose2},  (\state_{2},c_{1}) \mapsto \state_{n+1}\hdots,  (\state_{n},c_{n}) \mapsto \state_{{n\choose2}-1}\}$
\State $\vdots$
\State $\EdgeSet_{n-1} \leftarrow \{(\state_{2^n - (n+2)},c_n ) \mapsto \state_{2^n - 1}, \hdots,  (\state_{2^n - 2},c_1) \mapsto \state_{2^n - 1}\}$
\State $\EdgeSet_k \leftarrow \EdgeSet_1\cup\hdots\cup\EdgeSet_n$
\State Add $\EdgeSet_k$ to $\EdgeSet$
\State Add $\{(\state_{j},c_1)\mapsto AG_k.\state_{1}, \hdots, (\state_{j},c_n)\mapsto AG_k.\state_{n}\}$ to $\EdgeSet$
\State Add $(AG_k.\state_{2^n - 1},\actionitem) \mapsto \state_{k}$ to $\EdgeSet$
\ElsIf{$\operator = \SAND$}
\For{$c_i$ in $\Kid(t)$}
\State Add $\state_{j+i}$ to $\QStateSet$ and $(\state_{j+i-1}, c_i) \mapsto \state_{j+i}$ to $\EdgeSet$
\EndFor
\State Add $(\state_{j+n}, \actionitem) \mapsto \state_{k})$ to $\EdgeSet$
\EndIf
\State $\PStateSet \leftarrow \PStateSet\cup\QStateSet$
\State $\QStateSet \leftarrow \{\}$
\end{algorithmic}
\end{algorithm}

The algorithm starts with an attack graph containing two states, $\state_0$ and $\state_s$, the single starting and success states respectively. These two states are our initial $\PStateSet$ states. There is a single edge between these two states, which is defined as equivalent to the overall root of the attack tree. Intuitively, the meaning behind this simple 2-state attack graph is that accomplishing the overall goal represented by an attack tree (the root), we move from an initial state to a success state. However, accomplishing the overall goal is hardly a single step in and of itself, if it were the attack tree would be a single node as in Figure~\ref{fig:sig_node_ex}. By expanding the edge between $\state_0$ and $\state_s$ in the initial attack graph, we can add the detailed information of the subcomponents of the overall goal to the attack graph. As such, while processing the attack tree components in the radical dictionary, we add additional states and state transformations to the attack graph, until our attack graph fully represents the original attack tree.

The sequential construction of the attack tree follows from this intuition. We check the edges in the attack graph to see if they are a key for a radical in the radical dictionary. Once we find an edge that is a key to a radical in the radical dictionary, we remove this edge from the attack graph, and replace it with the relevant defined mapping from Section~\ref{sec:mapping}. The procedure of adding a radical to the attack graph is presented in Procedure~\ref{alg:RepAGEdge}. This expansion will cause new $\QStateSet$ states to be generated, which will then become $\PStateSet$ states for other radicals. Additionally, these mappings will cause the keys to other elements of the radical dictionary to be added to the attack graph, which subsequently allow for the addition of every radical in the radical dictionary to be added to the attack graph. After a radical is added to the attack graph, it is removed from the radical dictionary. Once the radical dictionary is empty, all the radicals from the attack tree has been added to the attack graph, and the attack graph is thus a complete transformation from the original attack tree, representing similar information.

\section{Discussion}
\label{sec:discussion}

\paragraph{Evaluation on the running example.}

If we return to the example found in Section \ref{sec:example}, we can see that the application of the algorithm as described in Figure~\ref{fig:AttackTree} will exactly result in the attack graph shown in Figure~\ref{fig:AttackGraph}. For every attack vector represented in the attack tree, the same attack vector is represented similarly in the attack graph, only with the introduction of states. These states are not provided with specific meaning outside of numbering as our specific definition; however, the ability to assign further meaning to these states is not excluded by our methodology.

The running example shows that the approach can create attack graphs from attack trees.  We now discuss the main benefits of our methodology, how can it be extended, and what measures have we taken to address the state explosion problem. For the lack of space, semantical aspects of the \OR\ decomposition and how does our approach relate to SP-graph semantics, the main semantics used for \SAND-attack trees~\cite{jhawarAttackTreesSequential2015}, are discussed in the Appendix~\ref{app:semantical}.

\paragraph{Benefits of our approach.}
One of the immediate benefits of our technique is that now it is possible to generate attack graphs from all human-designed attack trees from the literature and attack tree libraries. Indeed, most attack trees are created by humans, but attack graphs for any interesting system are generated automatically due to the state explosion problem, as discussed by Valmari~\cite{valmariStateExplosionProblem1998}. Thus by enabling the transformation from a human-generated attack tree to a transformed attack graph, we create an attack graph structure for a scenario of interest that potentially can be enhanced with the methods and tools created specifically for attack graphs (e.g., transformation into Bayesian networks~\cite{haqueEvolutionaryApproachAttack2017}). Moreover, typically attack trees use more abstract attack scenarios than attack graphs. Thus, it would be interesting to compare automatically generated attack graphs with the transformed attack graphs, to understand which attacks have been missed in both cases.

\paragraph{Future applications.}
Another advantage of enabling a direct transformation between attack trees and attack graphs is that new, automated generation schemes for attack graphs could become possible. As we mentioned earlier, significant research has already been performed into novel generation schemes for attack trees. A major branch of current attack tree research is into generation schema, particularly focusing on automated machine generation. Therefore, one advantage of being able to transform attack trees into attack graphs is the ability to use these novel generation methods to generate attack graphs.


While the transformation discussed here is in a single direction, as we discussed in Section~\ref{sec:defs}, we deliberately use the same definition of attack graphs as Pinchinat \et\ in their work on the ATSyRa algorithm which transforms attack graphs into attack trees \cite{pinchinatATSyRaIntegratedEnvironment2016}. As such, the attack graphs generated from our algorithm could be used as input into the ATSyRa algorithm to return them to attack trees.

This potentially enables the ability to convert an attack tree into an attack graph, the manipulation of the attack graph, and then conversion back into an attack tree. Furthermore, we now have a greater means to study potential edge cases within these models, for instance, the introduction of cycles within an attack graph, or the removal of a given state or specific attack vector.


Finally, thus far we have only concerned ourselves with a simple SAND attack tree \cite{jhawarAttackTreesSequential2015}, but further expansion of attack trees are available. For instance, we could expand this transformation to use an attack-defense tree (AD Tree) as an input, and then created an expanded attack graph representing the defensive actions represented in the attack tree. Expansion of the transformation methodology in this vein would similarly apply to the attack graph to attack tree ATSyRa transformation \cite{pinchinatSynthesisAttackTrees2014}.

\paragraph{Addressing state explosion problem.}

As pointed out by Antti Valmari, a major consideration in any state-based model will be the state explosion problem \cite{valmariStateExplosionProblem1998}. We thus want to define a sufficiently robust transformation, that results in an attack graph that contains enough information (\textit{i.e.} states) without containing too many states.

Our primary means of achieving a reduction in the state is to eliminate backtracking or partial attack paths. By maintaining the monotonicity in  generated attack graphs, we can prevent a factorial number of states from being generated. Back to the example in Figure~\ref{fig:AttackGraph}, we have no state such that the same action has been taken multiple times. There is no state $\state_i$ such that a path from $\state_0$ to $\state_i$ has two state transitions $\actionitem_i$ and $\actionitem_j$ where $i = j$. If we were to allow partial pathing, the attack graph couple potentially not terminate, or would have significant cycling.

Another mitigation strategy we take is the understanding that the \OR\ operator functions as more like an \texttt{X}\OR\ operator. 
We can see this in action in Figure~\ref{fig:AttackGraph}, where there is no state $\state_i$ such that a path from $\state_0$ to $\state_i$ has the state transitions \texttt{C} and \texttt{D}. If we were to understand that the \OR\ operator was not an \texttt{X}\OR, then we would have to add states to attack graphs to allow for each vector that includes multiple components from the \OR\ radical.
This directly increases the number of states for each child, given $n$ children, from $n$ to $n!$. We further discuss this choice in Appendix~\ref{app:semantical}.  

\section{Conclusions and Future Work}
\label{sec:conclusion}
In this work, we have laid out a structure of a transformation that will take an attack tree as input and return an attack graph. This style of transformation already exists in literature, albeit in the opposite direction as ours. We have evaluated our transformation approach on a case study.

In terms of expansion of our work, firstly we would wish to define a transformation of Attack-Defense Trees (AD Trees). Such a transformation would likely require an alternative definition of attack graphs; however, it may be possible to define a transformation such that we only use the definition of attack graphs that we use here. Furthermore, we would like to attempt to define a more efficient state creation schema, which would make the transformation algorithm more space and time efficient. 


\bibliographystyle{splncs04}
\bibliography{bibliography}

\appendix

\section{The $n$-\AND\ Radical}\label{app:sand}

\begin{equation}
\begin{aligned}\label{eq:n_AND}
\PStateSet &= \{\state_{0},\state_{s}\} \\
\QStateSet &= \{\state_{1},\hdots,\state_{n},\state_{n+1}, \hdots, \state_{2^n-1}\}\\
\EdgeSet_1 &=  \{(\state_{0},\text{Action 1}) \mapsto \state_{1}, \hdots,  (\state_{0},\text{Action n}) \mapsto \state_{n}\}\\
\EdgeSet_2 &= \{(\state_{1},\text{Action 2}) \mapsto \state_{n+1},\hdots, (\state_{n},\text{Action n-1})\mapsto\state_{n\choose2},\\
        &\text{ }\text{ }\text{ }\text{ }\text{ }\text{ }
           (\state_{1},\text{Action 1})\mapsto\state_{n+1}\hdots,  (\state_{n},\text{Action n}) \mapsto \state_{{n\choose2}-1}\}\\
\vdots \\
\EdgeSet_{n} &=\{(\state_{2^n - (n+2)}\text{Action n})\mapsto\state_{2^n - 1}), \hdots,  (\state_{2^n - 2},\text{Action 1})\mapsto \state_{2^n - 1}\}\\
\EdgeSet_{n+1} &=\{(\state_{2^n - 1},\text{Higher Action}) \mapsto \state_{s}\}\\
AG &= \left(\PStateSet\cup\QStateSet, \EdgeMap{\EdgeSet_1\cup\hdots\cup\EdgeSet_n}, \{\state_{0}\}, \{\state_{s}\}\right)
\end{aligned}
\end{equation}

We can see in Figure \ref{fig:AND_AG_rad} the creation of a lattice structure in the attack graph; this structure allows for all possible application orders for actions in an \AND\ radical. The $k^\text{th}$ row of the lattice represents the application of $k$ unique actions. As each subsequent row represents the application of one additional unique action, then we know there will be $n$ rows in the lattice; the final row will consist of a single node that is the application of all actions in the \AND\ radical. Selecting $k$ unique actions out of $n$ possible actions is the same as picking an unordered group of $k$ from $n$, which is the mathematical operator $n\choose k$. As every row will have $n\choose k$ possible states, the total number of states resolves to:
\[
\sum_{k=1}^{n} {n\choose k} = 2^n - 1
\]


\begin{figure}[t!]
\begin{subfigure}[t]{.5\textwidth}
\scriptsize
  \centering
  \scalebox{.8}{
  \begin{forest}
    for tree={
      draw,
      minimum height=1cm,
      anchor=parent,
      align=center,
      child anchor=parent,
      edge=<-
    },
    adnode/.style={rounded rectangle,},
      [{Higher Action}, adnode, angle below
        [{Action 1}, adnode]
        [{$\hdots$}, adnode]
        [{Action $n$}, adnode]
      ]
  \end{forest}

}
\label{fig:n_AND_AT_rad}
\caption{Attack Tree}
\end{subfigure}
\begin{subfigure}[t]{.5\textwidth}
\centering
\scalebox{.8}{
\scriptsize
  \centering
  \begin{tikzpicture}[->,>=stealth',shorten >=1pt,auto, node distance=2.2cm,
                      semithick]
    \tikzstyle{every state}=[fill=white,draw=black,text=black]

    \node[state,dashed] (0) [below left of=0] {$\state_{0}$};
    \node[state]         (1) [below left of =0] {$\state_{1}$};
    \node[state]         (3) [below right of=0] {$\state_{n}$};
    \node[state]         (2) at ($(1)!0.5!(3)$) {$\state_{k}$};
    \node[state]         (4) [below left  of=1] {$\state_{n+1}$};
    \node[state]         (6) [below right of=3] {$\state_{n\choose{2}}$};
    \node[state]         (5) at ($(4)!0.5!(6)$) {$\state_{n+k}$};
    \node         (Ml) [below=.3cm of 4 ] {};
    \node         (Mr) [below= .3cm of 6 ] {};
    \node         (M) [below= .3cm of 5 ] {\Huge$\hdots$};
    \node[state]         (7) [below=.5cm of Ml] {$\state_{2^n - (n+2)}$};
    \node[state]         (9) [below= .5cm of Mr] {$\state_{2^n-2}$};
    \node[state]         (8) at ($(7)!0.5!(9)$) {$\state_{2^n - (n+2) + k}$};
    \node[state]         (10) [below = .5 cm of 8] {$\state_{2^n - 1}$};
    \node[state,dashed]         (G) [below = .5 cm of 10 ] {$\state_{s}$};

    \path (0) edge              node [left]{Action 1} (1)
              edge              node {$\hdots$} (2)
              edge              node [right]{Action $n$} (3)
          (1) edge              node [left]{$\hdots$} (4)
              edge              node {$\hdots$} (5)
          (2) edge              node [sloped,pos=.6] {Action 1} (4)
              edge              node [sloped,pos=.6] {Action $n$} (6)
          (3) edge              node {$\hdots$} (5)
              edge              node [right]{$\hdots$} (6)
          (7) edge              node [sloped,]{Action $n$} (10)
          (8) edge              node {$\hdots$} (10)
          (9) edge              node [sloped,]{Action 1} (10)
          (10)edge              node {Higher Action} (G);

  \end{tikzpicture}
}
\caption{Attack Graph}
\label{fig:n_AND_AG_rad}
\end{subfigure}
\caption{$n-$\AND\ Radical}
\label{fig:n_AND}
\end{figure}

\section{Semantical Considerations}\label{app:semantical}
\subsubsection{\texttt{X}\OR-like \OR.}
\label{ssec:xor}

Our interpretation of the \OR\ refinement resembles more an exclusive-\OR. To illustrate this point, consider again the \OR\ radical mapping shown in Figure~\ref{fig:OR_AG_rad}, in this mapping, we assume that only one of the \OR\ actions will be completed. While a common understanding of an \OR\ operator is that the output is true if one input is true, in our case, we are explicitly excluding the possibility of multiple actions happening in a single \OR\ radical. Our primary reason for this is to reduce the number of states in the generated attack graph, which is further discussed in the next sub-section. It would be possible to allow a true \OR, where multiple children of an \OR\ radical can be components of a successful attack vector. These attack vectors would have redundant components, as including more than one subcomponent of an \OR\ would introduce redundancy.

\subsubsection{Relationship with SP-graph semantics}
\label{ssec:sp-graph}

SP-graphs are one method of semantically representing attack trees \cite{jhawarAttackTreesSequential2015}. If we consider the following small attack graph and its SP-graph semantics, we can see why our approach is fundamentally different than SP-graph semantics.

While network expressions of SP graphs and attack graphs might look similar, these structures are fundamentally different. SP graphs do not have a concept of states while states are central to attack graphs. Additionally, the common understanding of SP graphs focuses entirely on leaf nodes, with the understanding that intermediate nodes do not carry information and are primarily placeholders \cite{mauwAttackTrees2006, jhawarAttackTreesSequential2015}. Manually created attack trees are most likely to have intermediate nodes that carry meaning, and as such, should be included in any semantic representation of that attack tree. Our generated attack graphs assume that intermediate nodes carry meaning, which SP graphs assume that intermediate nodes are purely functional without independent meaning.

Given SP graphs are a valid semantic representation of attack trees, our attack graph generation schema requires an attack tree as input, and the visual similarities between these representations, it follows that attack graphs could potentially be generated from SP graphs directly. However, when examining our transformation algorithm, the most computationally intensive component is the generation and connection of the states in attack graphs. And as we just discussed, SP graphs do not have states. As such, an algorithm using SP graphs as input would still have a computationally expensive state generation and connection step, and would not be any faster or more efficient.

\begin{equation}\label{eq:SP-Counter-Example}
  \left\{\xrightarrow{\text{A}}||\xrightarrow{\text{C}},
  \xrightarrow{\text{A}}||\xrightarrow{\text{D}}\right\}
\end{equation}

\end{document}